\documentclass[%
 reprint,
 superscriptaddress,
 amsmath,amssymb,
 aps,
floatfix,
]{revtex4-1}

\usepackage{graphicx}
\usepackage{dcolumn}
\usepackage{bm}
\usepackage{braket}
\usepackage{color}%
\usepackage{url}%

\newcommand{\affA}{%
\affiliation{
     National Institute of Information and Communications Technology
     (NICT), \\
     4-2-1 Nukui-kitamachi, Koganei, Tokyo 184-8795, Japan}
     }
\newcommand{\affB}{%
\affiliation{
     PRESTO, Japan Science and Technology Agency, Honcho, Saitama 223-0012, Japan
    }
}

\begin{document}


\title{Generation of vacuum ultraviolet radiation by intracavity high-harmonic generation toward state detection of single trapped ions}

\date{\today}

\author{Kentaro Wakui}
\affA%
\author{Kazuhiro Hayasaka}
\affA%
\author{Tetsuya Ido}
\affA\affB%
\begin{abstract}
VUV radiation around 159\,nm is obtained toward direct excitation of a single trapped $^{115}$In$^{+}$ ion. 
An efficient fluoride-based VUV output-coupler is employed for intracavity high-harmonic generation of a Ti:S oscillator. 
Using this coupler, where we measured its reflectance to be about 90\,$\%$, 
an average power reaching $6.4\,\mu$W is coupled out from a modest fundamental power of 650\,mW. 
When a single comb component out of $1.9\times10^{5}$ teeth is resonant to the atomic transition, 
hundreds of fluorescence photons per second will be detectable under a realistic condition.
\end{abstract}

\maketitle

\section{Introduction}
\label{introduction}

\begin{figure}
\resizebox{0.5\textwidth}{!}{%
  \includegraphics{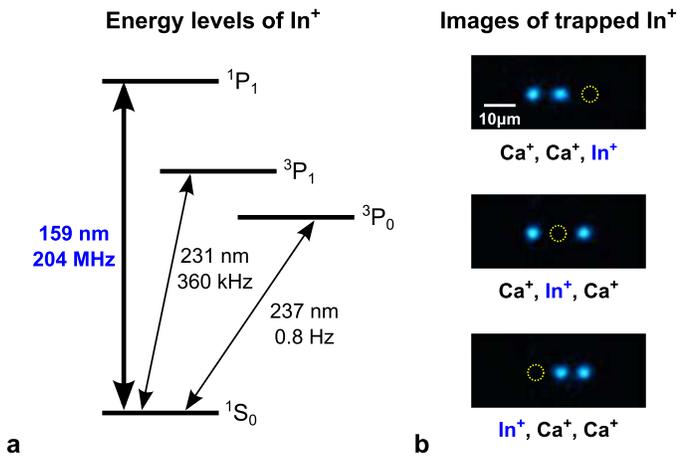}
}
\caption{\textbf{(a)} Relevant energy levels and transitions of In$^+$. 
\textbf{(b)} Images of In$^+$ sympathetically cooled with Ca$^+$ 
observed by resonant fluorescence of Ca$^+$ at 397\,nm. 
The In$^+$ is identified as the dark site in the ion chains.}
\label{indium_levels}       
\end{figure}

Optical clocks based on trapped single ions and neutral atoms in optical lattices 
have reached frequency uncertainties in the order of  $10^{-18}$ \cite{Chou10,Bloom14,Ushijima14}. 
This level of accuracy brings strong impact to applications in science and technology, 
such as precise determination of physical constants, tests of fundamental physics theories, 
network synchronization, geodesy, and redefinition of the second.
Many of the optical clocks use atoms and ions with two outer electrons
as the optical frequency reference according to the original proposal by Dehmelt \cite{Dehmelt82}. 
The idea of single ion clocks has been extended to optical lattice clocks
where multiple neutral atoms are confined in the Lamb-Dicke regime \cite{Katori03}.
Energy levels of such atomic species (B$^+$, Al$^+$, Ga$^+$, In$^+$, Tl$^+$, Sr, Yb, Hg) 
is shown in Fig. \ref{indium_levels}a with In$^+$ as an example. 
Their weakly allowed ${}^1\rm{S}_0$--${}^3\rm{P}_0$ transitions with linewidth 
smaller than 1\,Hz are used as clock transitions. 
The clock transitions are observed using the ${}^1\rm{S}_0$--${}^1\rm{P}_1$ transitions as follows:
Excitation of the ${}^1\rm{S}_0$--${}^1\rm{P}_1$ transition 
after the interrogation of an atom with clock lasers results in a lot of fluorescent photons 
when the clock laser is off-resonant from the ${}^1\rm{S}_0$--${}^3\rm{P}_0$ transition,  
and results in absence of photons when it is resonant.
This method called ``electron shelving" is relatively easily implemented to lattice clocks
since the ${}^1\rm{S}_0$--${}^1\rm{P}_1$ transition lies in visible wavelength.
In contrast, application to the ions is intractable, 
because the ${}^1\rm{S}_0$--${}^1\rm{P}_1$ transitions locate in the vacuum ultraviolet (VUV) region 
as is exemplified in Fig. \ref{indium_levels}a, 
where generation of coherent radiation is challenging.
The VUV radiation is also expected to laser-cool the ion, and this demands the VUV radiation to be continuous-wave (CW).

The solution to overcome this difficulty has been provided by quantum logic spectroscopy (QLS), 
in which another ion with a convenient energy level structure is trapped simultaneously 
and serves as a coolant ion as well as a quantum state probe \cite{Schmidt05}. 
By this way the Al$^+$ clock has been realized without need for VUV radiation 
to excite the ${}^1\rm{S}_0$--${}^1\rm{P}_1$ transition at 167\,nm.
The smallest inaccuracy of single-ion clocks has been reported with the Al$^+$ clock \cite{Chou10}, 
but the implementation of QLS seems technically demanding
since no other demonstrations are so far reported.

Here, we propose an alternative approach for the ion clock, 
in which VUV radiation is generated via high-harmonic generation (HHG) \cite{Brabec00,Atto_Sci_07},
and is used only for quantum state detection of single ions.
Cooling of the ions is supplied by sympathetic cooling, 
and therefore, a single-mode CW radiation is not required.
Detection of single ions might be suffice with a quasi-CW radiation 
which is actually a singled-out mode in ``VUV combs" generated via intracavity HHG \cite{Jones05,Gohle05}.
Sympathetically cooled ions are relatively easily prepared \cite{Hayasaka12,Herschbach12}, 
and an example of such ion arrays including an In$^+$ is shown in Fig. \ref{indium_levels}b.
Besides this main advantage of simpleness, 
our method has another advantage of scaling to multi-ion systems for improved stability \cite{Herschbach12}.
Although the current protocol of QLS is limited to systems with one target ion, 
the state detection using VUV radiation is applicable to systems with multiple target ions.
Such a quasi-CW beam in the VUV region might be used as a clock laser that interrogates a thorium nucleus 
for realizing a more accurate optical clock \cite{Peik03,Beck07,Rellergert10,Campbell11,HS13}. 
Several methods which might be applicable to direct excitation of VUV transitions 
have been developed as follows.
Triply-resonant four-wave mixing is implemented to generate 121\,nm light \cite{Kolbe12}.  
KBBF crystals are used to demonstrate generation of 156\,nm radiation by sum-frequency mixing \cite{Kanai04} 
and generation of 153 nm radiation via frequency conversion \cite{Nomura11}. 
Raman-resonant four-wave mixing in parahydrogen has been proposed more recently 
to generate 120--200\,nm VUV radiation \cite{Zheng14}.

In section \ref{feasibility_study} we discuss the feasibility of detecting the ${}^1\rm{S}_0$--${}^1\rm{P}_1$ 
fluorescent photons at 159\,nm from a single In$^+$ ion excited with a quasi-CW beam obtained via intracavity HHG.
Section \ref{experiment} describes experiments to generate the VUV radiation 
as the fifth harmonic of femtosecond (fs) frequency combs peaked at 795\,nm.
We conclude this work and discuss the future prospects in section \ref{conclusion}.

\section{Single ion detection with a quasi-CW VUV beam}
\label{feasibility_study}

\begin{figure}
\resizebox{0.5\textwidth}{!}{%
  \includegraphics{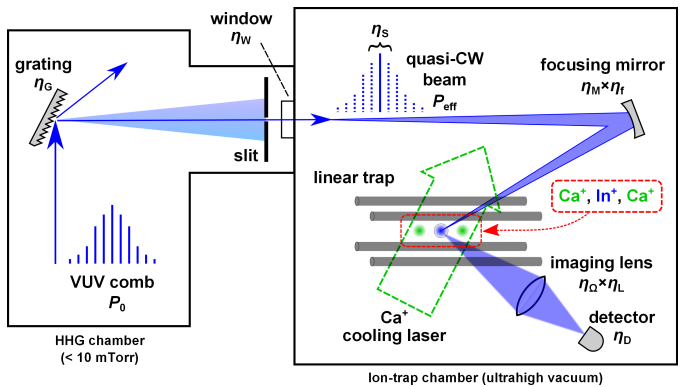}
}
\caption{Setup for observing a sympathetically cooled In$^+$ by direct excitation 
using a quasi-CW beam generated from the VUV comb.}
\label{detection_setup}       
\end{figure}

In this section we discuss on the possibility of detecting the quantum state of a single trapped ion 
by direct excitation using a quasi-CW beam generated as VUV combs.
Our target is the $^{115}$In$^+$, 
which is one of the candidates for single-ion optical clocks included in Dehmelt's original proposal \cite{Dehmelt82}. 
Recent proposals on the In$^+$ optical clocks estimate fractional frequency inaccuracy of 10$^{-18}$ level 
owing to its small black body radiation shift \cite{Hayasaka12,Herschbach12}. 
Previously, implementation of the In$^+$ clock has been investigated 
using relatively weak ${}^1 \rm{S}_0$--${}^3 \rm{P}_1$ transition at 230\,nm for detection, 
but its performance was limited to 10$^{-13}$ level \cite{vonZanthier00,Wang07}.
If the direct excitation of the ${}^1 \rm{S}_0$--${}^1 \rm{P}_1$ transition at 159\,nm is possible, 
roughly 500 times more fluorescent photons are emitted from a single In$^+$, 
and it makes much faster quantum state detection than that with the ${}^1 \rm{S}_0$--${}^3 \rm{P}_1$ transition.
Combined with robust trapping of the ions by sympathetic cooling, 
the VUV detection scheme will enable an easier implementation of the optical clock.
The same method might be applicable not only to a multi-In$^+$ clock, 
but also to a single Al$^+$ clock as well as to multi-Al$^+$ clock, 
where the ${}^1 \rm{S}_0$--${}^1 \rm{P}_1$ transition lies at 167\,nm. 

We assume a setup depicted in Fig. \ref{detection_setup} 
for direct excitation of a single In${}^+$ with a VUV beam generated from HHG. 
The whole optical path between the VUV source and the detector is evacuated to avoid absorption by air: 
The VUV source and the ion trap will be deployed in separate vacuum chambers 
for their different vacuum levels, 
and will possibly be connected through a high-transmittance fluoride window.
We focus on the case of single ion in this paper, 
but extension to multiple ions might be possible by upgrading the optics for simultaneous illumination of ions.   
The VUV comb is incident on a grating, 
and modes around the ${}^1 \rm{S}_0$--${}^1 \rm{P}_1$ transition are chosen by a slit to avoid detrimental influence 
by excess modes such as increased background photons due to scattering, 
and charging of insulating materials due to photoelectric effect.
The quasi-CW beam generated in this way is focused onto the In$^+$ sympathetically cooled with two Ca$^+$. 
Fluorescent photons emitted from the In$^+$ are imaged onto the detector using a single lens. 
Quantum state detection of the single ions is enabled 
by sufficient number of fluorescent photons detected with a photomultiplier tube (PMT)
or a image-intensified charge-coupled-device (CCD) camera.

In order to assess the feasibility of the VUV detection method,  
it is crucial to know the total average power $P_0$\,[W] of the VUV comb  
required for detecting $n_0$\,[cps] (counts per second) fluorescent photons with the detector.
Rough estimation is made on $P_0$ by considering the steps of the direct excitation of single ion 
consisting of generation of  the quasi-CW beam focused to the ion, 
fluorescent photon scattering by the single ion and collection and detection by the detector.
The VUV comb is incident on the grating, 
and only the mode interacting with the ion is transmitted through the slit, 
and is focused with a curved mirror onto the ion.
After this step, the quasi-CW power $P_{\rm{eff}}$ which effectively contributes to excitation of the ion,  
is given by
\begin{equation}
P_{\rm{eff}}
=
\eta_{\mathrm{prop}} \times P_0.
\end{equation}
Here 
$
\eta_{\mathrm{prop}} 
= 
\eta_{\mathrm{G}} \eta_{\mathrm{W}} \eta_{\mathrm{S}} \eta_{\mathrm{M}} \eta_{\mathrm{f}}
$
is the total propagation efficiency to the ion,
in which
$\eta_{\mathrm{G}}$, $\eta_{\mathrm{W}}$, $\eta_{\mathrm{S}}$, $\eta_{\mathrm{M}}$ and $\eta_{\mathrm{f}}$ denote, 
respectively, grating efficiency, window transmittance, power ratio of the target comb-mode relative to the whole spectrum, 
reflectance of the focusing mirror and focusing efficiency to the spot with a radius of $r_0$.
The ion emits fluorescent photons with a rate of
\begin{equation}
n_{\rm{fl}} 
=
\frac{P_{\rm{eff}}}{A_{\rm{spot}}} \frac{\Gamma}{2I_{\rm{s}}},
\end{equation}
where $\Gamma$ is the spontaneous emission rate and $I_{\rm{s}}$ is the saturation intensity, 
respectively, of the ${}^1\rm{S}_0$--${}^1 \rm{P}_1$ transition. 
$A_{\rm{spot}} = \pi r_0^2 $ is the spot size.
The photons are imaged onto a detector with a quantum efficiency of $\eta_{\rm{D}}$ 
by a single lens which spans a solid angle covering $\eta_{\rm{\Omega}}$ of the whole solid angle, 
and has a transmittance of  $\eta_{\rm{L}}$.
Then the count rate 
$
n_0 = \eta_{\rm{det}} \times n_{\rm{fl}}
$
\,[cps] is obtained by the detector,
where 
$\eta_{\rm{det}} = \eta_{\rm{\Omega}} \eta_{\rm{L}} \eta_{\rm{D}}$ 
is an overall efficiency in the detection side.
Thus the photon counting rate per the average VUV-comb power 
can be estimated by the following expression: 
\begin{equation}
\frac{n_0}{P_0} = \frac{ \eta_{\rm{prop}} \times \eta_{\rm{det}} }{A_{\rm{spot}}} \frac{\Gamma}{2I_S}.
\end{equation}

\begingroup
\renewcommand{\arraystretch}{1.3}
\begin{table}
\begin{center}
\begin{tabular}{lcl} \hline
\multicolumn{1}{c}{description} & symbol & estimation \\ \hline
spontaneous emission rate & $\Gamma$ & 2$\pi \times$204\,MHz \\
saturation intensity & $I_{\rm{s}}$ & 6.67\,W/cm$^2$ \\ 
spot radius & $r_0$ & 0.5\,$\mu$m \\
grating efficiency & $\eta_{\rm{G}}$ & 0.34  \\
window transmittance & $\eta_{\rm{W}}$ & 0.85  \\
spectrum occupancy  & $\eta_{\rm{S}}$ & 5.3$\times$10$^{-6}$ \\
mirror reflectance & $\eta_{\rm{M}}$ & 0.95 \\
focusing efficiency & $\eta_{\rm{f}}$  & 0.85 \\ 
photon collection efficiency & $\eta_{\rm{\Omega}}$ & 0.045 \\ 
imaging lens transmission & $\eta_{\rm{L}}$ & 0.85  \\ 
detector quantum efficiency & $\eta_{\rm{D}}$ & 0.15  \\ \hline
\end{tabular}
\caption{Assumption on numbers for estimating the total average power. 
The details are described in the text.}
\label{tab:eta}     
\end{center}
\end{table}
\endgroup

Now we proceed with estimation on $P_0$ 
required for the single ion detection with realistic experimental parameters.
The spontaneous emission rate of the ${}^1\rm{S}_0$--${}^1 \rm{P}_1$ 
transition is lately revised to be $\Gamma = 2\pi\times204$\,[MHz] \cite{NIST_ASD}. 
The saturation intensity of the transition is calculated from 
$\displaystyle I_{\rm{s}} = \frac{\pi hc \Gamma}{3\lambda^3}$, 
where $h = 6.63\times 10^{-34}$\,Js is the Planck constant 
and $c$ is the speed of light.
Using the above value of $\Gamma$ and wavelength $\lambda$ = 158.6\,nm, 
the saturation intensity is $I_{\rm{s}}$ = 6.67\,W/cm$^{2}$.
The advantage of sympathetically cooled ions is that 
even the invisible ion can be localized within a known region much smaller than micrometer scale. 
Localization of a Ca${}^+$ Doppler-cooled in a conventional linear trap 
after a routine procedure of micromotion compensation 
was measured to be as small as 16\,nm (full width at half maximum) 
along the trap axis direction \cite{Guthohrlein01}. 
The wave packet size of the In${}^+$ cooled by this Ca${}^+$ is 
estimated $\sqrt{115/40}$ times larger 
due to dependence of the axial secular frequency on mass ($\propto 1/\sqrt{m}$), 
but the localization is still in the order of tens of nanometers. 
This allows targeting the ion with the smallest spot size possible with focusing optics.
Here we consider a simple case  
in which a collimated Gaussian beam is tightly focused by a concave mirror 
(denoted as ``focusing mirror" in Fig. \ref{detection_setup}). 
When we take a mirror diameter $D = 25\,$mm and a radius of curvature $R = 50$\,mm as an example,
the effective diameter of the focused Gaussian spot is estimated to be 
$\displaystyle d \approx \frac{R\lambda}{D} \approx 0.32\,\mu$m. 
Thus we assume that the VUV radiation is focusable to the spot radius e.g. $r_0$ = 0.5\,$\mu$m, 
which is more than three times larger than $d/2$. 
We also estimated the focusing efficiency $\eta_{\rm{f}}$ = 0.85, 
which denotes the reduction of intensity from the ideal focusing.
Later we discuss on derivation of $\eta_{\rm{f}}$ in some detail.
Grating efficiency of $\eta_{\rm{G}}$ = 0.34 was realized at 160\,nm \cite{Caruso81}.
We estimated that the intensity ratio of the target comb-mode to the whole spectrum, namely spectrum occupancy, 
is $\eta_{\rm{S}}$ = 5.3$\times 10^{-6}$,
which is an important issue and is discussed in detail later.
Mirror reflectance $\eta_{\rm{M}}$ is estimated to be 95\,$\%$,
because such a high-reflection coating is available at the wavelengths longer than 150\,nm \cite{VUV_spec_book}.
For collection of fluorescent photons, 
we assume a use of a lens with a focal length of $f$ = 30\,mm in 25\,mm diameter made of MgF$_2$.
This gives photon collection efficiency of $\eta_{\rm{\Omega}}$ = 0.045.
A typical transmittance of $\approx85\,\%$ was realized with a 3-mm-thick MgF$_2$ \cite{VUV_spec_book},
and thus we use a lens transmission efficiency of $\eta_{\rm{L}} = 0.85$.
Note this value is also used for the window transmittance $\eta_{\rm{W}}$.
The quantum efficiency of our image intensified CCD camera (PI-MAX 3, Princeton Instruments) is 0.15 at 159\,nm,
which is similar to the maximum efficiency of PMTs. 
Therefore we take this value as the detector efficiency $\eta_{\rm{D}}$.

Next we estimate the focusing efficiency $\eta_{\rm{f}}$ in detail.
We take into account of aberration caused by the focusing mirror
which reflects the VUV radiation originally generated as a Gaussian beam,
because the accuracy of mirror surface is closely related to the beam focusing efficiency.
Surface accuracy is usually measured at 633\,nm (typically using a helium-neon laser), 
and $\lambda/10$ ($ = 63.3$\,nm) can be easily achievable for the mirror diameter $D\sim25$\,mm. 
The accuracy of 63.3\,nm corresponds to phase aberration of $A = 2.5$\,rad at 159\,nm.
Then one can evaluate the Strehl ratio of a Gaussian beam,
$S = \exp(-\sigma_{\rm{\Phi}}^2)$,
which represents the ratio of the central irradiance of point-spread function 
before and after being affected by aberrations \cite{Mahajan05}.
$\sigma_{\rm{\Phi}}^2$ is the variance of the phase aberration in a Gaussian beam.
Due to the fact that surface roughness of a polished mirror is usually below a few angstroms, 
and that spatial frequency of surface irregularity is not high,
we consider monochromatic primary aberrations, 
e.g. spherical aberration ($\sigma_{\rm{\Phi}} = A/6.20$), coma ($\sigma_{\rm{\Phi}} = A/6.08$), 
and astigmatism ($\sigma_{\rm{\Phi}} = A/6.59$) \cite{Mahajan05}.
In each case, the Strehl ratio of around $85\,\%$ can be maintained.
In other words, roughly $85\,\%$ power in the reflected VUV beam 
is not blurred and will be correctly focused 
even with aberration which may be caused by the practical surface accuracy. 
Note, further improvements of the Strehl ratio is possible by combining and balancing several aberrations \cite{Mahajan05}.
In contrast to the focusing mirror described above, 
the beam quality would not be affected as much by a surface irregularity of a dichroic beam splitter 
which should be placed in advance of the beam input in Fig. \ref{detection_setup} 
and separates fundamental and high-harmonic beams. 
As shown later in this paper, the beam diameter at the splitter is normally as small as 1\,mm. 
The corresponding primary aberrations caused by that small area 
is much smaller than that in the case of the concave focusing mirror ($D \sim 25$\,mm) 
because spherical aberration, coma, and astigmatism vary with the fourth, third, and second power, respectively, 
of radial distance from the beam axis. 
Thus the Strehl ratio for the splitter is estimated to be near unity. 
Note that surface accuracy of $\lambda/10$ is commercially available 
in substrates with the thickness and the diameter of 1\,mm and 5\,mm, respectively.

Finally we estimate the spectrum occupancy $\eta_{\rm{S}}$ as follows.
Nonlinear optical processes in the HHG can be categorized as
\textit{perturbative}, \textit{intermediate}, and \textit{strong-field} regimes
according to the strength of fundamental electric field interacting with atoms \cite{Brabec00}.
As we show in section \ref{intracavity_HHG},
the peak intensity $I_{\rm{peak}} $ of fundamental fs pulses in our setup
is estimated to be around $2\times10^{13}$ W/cm${}^2$,
which is presumably within the perturbative regime.
In fact, we calculated the scale parameters in the ref. \cite{Brabec00} as follows, 
$\displaystyle \alpha_{bb} = \frac{e E_a a_{\rm{B}}}{\hbar \Delta} \approx 1.1\times 10^{-3}$ 
and 
$\displaystyle \alpha_{bf} = \frac{e E_a a_{\rm{B}}}{\hbar \omega_0} \approx 4.5\times 10^{-3}$.
Here, 
$\displaystyle e = 1.6\times10^{-19}$\,C, 
$\displaystyle E_a = \sqrt{2 Z_0 I_{\rm{peak}}} \approx 1.3\times10^{8}$\,V/cm, 
and 
$\displaystyle a_{\rm{B}}=\frac{\hbar}{\sqrt{2 m W_{\rm{Xe}}}} \approx 5.6\times10^{-11}$\,m, 
denote, respectively, the elementary charge, electric field strength, 
and the Bohr radius of Xe atoms used as nonlinear media for HHG.
We also used 
$m = 9.11\times10^{-31}$\,kg, $Z_0 = 377$\,V/A, 
and $W_{\rm{Xe}} = 12.13$\,eV,
representing the electron rest mass, the impedance of vacuum, 
and the ionization potential of the Xe atoms \cite{NIST_ASD}, respectively.
$\displaystyle \hbar = \frac{h}{2\pi}$ is the reduced Planck constant and 
$\displaystyle \hbar \omega_0 
= 1.57$\,eV is a photon energy at 795\,nm.
$\displaystyle 1/\Delta = 1/|\omega_{159\rm{nm}}-\omega_{0}| \approx 0.1 $\,fs/rad 
is a response time of induced atomic dipole moment.
Thus we confirmed that the following condition is fulfilled as
$\alpha_{bb} \ll 1$
and
$\alpha_{bf} \ll 1$,
that is the definition of the perturbative regime \cite{Brabec00}.
This may allow us to describe electric field for the $n$th harmonic as 
$e_n(t) \propto \left[ e_{\rm{F}}(t) \right]^n$,
where $e_{\rm{F}}(t)$ denotes the electric field at fundamental wavelength.
If we assume a transform-limited pulse e.g. with a Gaussian temporal envelope 
whose width is characterized by $\sigma$,
simple Fourier transform analysis reveals that 
the spectral width for the $n$th harmonic in the frequency domain 
becomes $\sqrt{n}$ times broader than that for the fundamental:
$
\displaystyle e_n(t) \propto \exp \left( - \frac{n t^2}{2\sigma^2} \right) 
\rightarrow 
\displaystyle E_n(\omega) \propto \exp \left( - \frac{n \sigma^2 \omega^2}{2} \right).
$
As we show in section \ref{intracavity_HHG}, a spectral bandwidth of the fundamental fs pulses 
is $B_{\rm{F}} = 9.5$\,THz (20\,nm centered at 795\,nm).
This may yield $B_{\rm{5th}} = \sqrt{5} \, B_{\rm{F}} = 21.2$\,THz 
($159\pm0.9$\,nm) for the fifth harmonic.
$\displaystyle \eta_{\rm{S}} = (B_{\rm{5th}} / f_{\rm{rep}} )^{-1} \approx 5.3\times10^{-6}$
is obtained as an inverse of the number of VUV-comb teeth ($ \approx 1.9\times10^{5} $).
Here we introduced an experimental repetition rate of our fs frequency comb as $f_{\rm{rep}} = 111.5$\,MHz.

In summary, the photon counting rate per the average VUV-comb power, 
$n_0/P_0$, is estimated to be roughly 87\,cps/$\mu$W, using the above parameters summarized in table \ref{tab:eta}.
According to the previous report 
in which single In$^+$ is detected using the ${}^1 \rm{S}_0$--${}^3 \rm{P}_1$ transition at 230\,nm \cite{Wang07},
the quantum state detection was demonstrated with a photon counting rate of 500\,cps. 
This rate is obtained with $P_0 = 5.8\,\mu$W in our scheme. 
In the following section \ref{experiment}, 
we experimentally confirmed that this level of power can be generated using our existing setup.

\section{Intracavity high-harmonic generation using VUV-OC}
\label{experiment}
High-harmonic generation (HHG) has been intensively studied 
for generation of radiation whose wavelengths are shorter than 190\,nm VUV region,
in order to realize attosecond pulses enabling one to probe and control 
electron dynamics in the time domain \cite{Atto_Sci_07}.
The Ti-sapphire (Ti:S) based system is convenient for a tool of spectroscopy 
because it can generate various wavelength. 
The HHG process, however, requires very strong laser intensities ($>10^{13}$\,W/cm${}^2$). 
Though chirped pulse amplification of Ti:S oscillator has been widely used to realize such a condition, 
its repetition rate cannot easily exceed 20 kHz \cite{Brabec00}.
Alternatively, intracavity HHG using fs frequency combs 
has also realized such a strong intensity by using a passive fs enhancement cavity (fsEC),
while it can maintain a repetition rate of 100-MHz-level.
The intracavity HHG thus realized frequency combs in the VUV 
and extreme ultraviolet (XUV) region \cite{Jones05,Gohle05}.
More recently, direct frequency-comb spectroscopy using such VUV or XUV combs
have been implemented for probing transition frequencies 
in argon at 82\,nm and in neon at 63\,nm \cite{Cingoz12}, 
or in xenon at 147\,nm \cite{Ozawa13}.
The XUV combs may also be useful for testing bound-state quantum electrodynamics
in a single helium ion \cite{Herrmann09}.

Considering to apply the direct state detection scheme to an In${}^+$ clock 
being developed in our institute \cite{Hayasaka12}, 
we built a setup of an intracavity HHG and reported 
VUV-comb generation 
centered at 159\,nm \cite{Wakui11}.
But further power-scaling is needed in order to realize such a VUV detection 
of single trapped ions.
One of the straightforward ways to scale up the harmonic yield 
would be just to enlarge fundamental power of the fs frequency combs
\cite{Paul08,Ruehl10} for HHG.
The outcoupling method of high harmonics from the fsEC 
is also the key to scale up their yield.
Several methods were previously demonstrated using 
e.g. a bulk sapphire window as a Brewster plate \cite{Jones05,Gohle05}, 
a diffraction grating ruled onto a fsEC mirror surface \cite{Yost08}, 
a diffraction nanograting \cite{Yang11},
an anti-reflection-coated grazing incidence plate \cite{Pronin11},
a VIS/IR-XUV beamsplitter \cite{Pupeza11}, 
a Brewster plate made of MgO substrate with a dielectric coating for VUV \cite{Ozawa13}, 
and a pierced fsEC mirror \cite{Pupeza13}.

In this section, we report a novel efficient VUV output coupler (VUV-OC) for the fifth harmonic of Ti:S oscillator. 
The VUV-OC consisting of a thin SiO${}_2$ substrate and fluoride-multilayer coating,
works as a Brewster plate which can be inserted in the fsEC driven at a fundamental near-infrared (NIR) wavelength, 
and at the same time works as a high reflector at the VUV wavelength.
In fact, we demonstrated an efficient outcoupling of the VUV comb generated at around 159\,nm,
and obtained an average power of 6.4\,$\mu$W measured with a calibrated detector.
This VUV power is almost ten times larger than that obtained in our previous experiment \cite{Wakui11}.

\subsection{Development of fluoride-multilayer-coated Brewster plate (VUV-OC)}
\label{VUV-OC}

The design and performance of the VUV-OC is described here.
The substrate of our VUV-OC is a SiO${}_2$ glass plate 
with thickness of 100\,$\mu$m.
The Brewster angle of the substrate is $55^{\circ}$ at 795\,nm.
On one surface, we deposited a high-reflection coating at 159\,nm, 
which is a multilayer quarter-wave stack 
consisting of AlF${}_3$ and GdF${}_3$ for low- and high-refractive-index materials, respectively.
The VUV-OC reflectance $R_{\mathrm{OC}}^{\rm{VUV}}$ is designed to be $>80\,\%$ at 153--163 nm wavelengths
and $>90\,\%$ at 159\,nm for p-polarized VUV light. 
The VUV-OC was made by Sigma-Koki.
An average reflectance of the VUV-OC at NIR wavelengths from 760 to 840\,nm is
measured as $R_{\mathrm{OC}}^{\rm{NIR}} = 0.1150\pm0.005\,\%$.
This is mainly caused by the fluoride-coating layer, and may act as a major loss inside the fsEC.
Note, because our VUV-OC is thin, 
its group delay dispersion (GDD) is small enough in the above NIR range.

\begin{figure}
\resizebox{0.5\textwidth}{!}{%
  \includegraphics{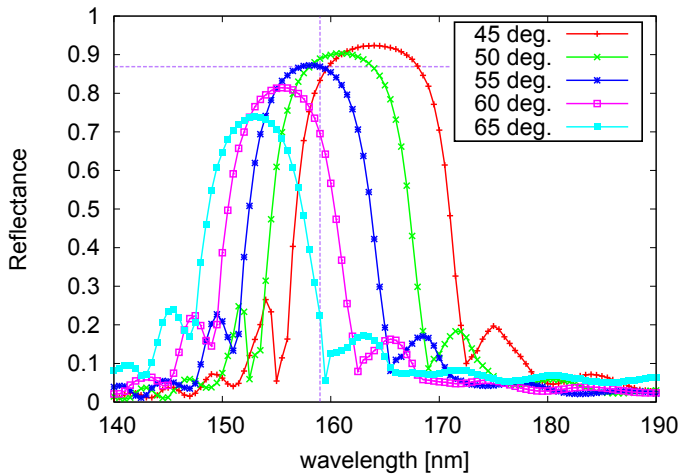}
}
\caption{VUV-OC reflectance measured at 140--190\,nm.
From left to right, 5 splined curves correspond to the incident angles of 
$65^{\circ}$(cyan), $60^{\circ}$(magenta), $55^{\circ}$(blue), 
$50^{\circ}$(green), and $45^{\circ}$(red).
For reference, 
the purple-dotted horizontal and vertical lines indicate 
reflectance of $0.869$ and wavelength of 159\,nm, respectively.}
\label{reflectance}       
\end{figure}

In order to evaluate the reflectance of VUV-OC for p-polarization in the VUV range,
we carried out reflectance measurement using a synchrotron light source 
at UVSOR facility (Okazaki, Japan) \cite{BL5B}.
First, incident VUV power from the storage ring is calibrated by a photodiode (AXUV100, IRD).
Then the reflection from the VUV-OC, which was placed on a rotation stage, 
was measured by the same photodiode.
Thus we measured the absolute reflectance shown in Fig. \ref{reflectance}
at several incident angles 
($45^{\circ}$, $50^{\circ}$, $55^{\circ}$, $60^{\circ}$, and $65^{\circ}$). 
Wavelength was changed with a monochromator in the beamline 
from 140 to 190\,nm with a wavelength step of 0.5\,nm.
At the Brewster angle of $55^{\circ}$, 
the reflectance of $86.9\pm0.4\%$ was obtained at 159\,nm, 
and $>80\,\%$ was achieved from 155 to 161.5\,nm.
We found that the maximum reflectance is higher 
at the smaller incident angles: 
$90.3\pm0.6\,\%$ at 161\,nm ($50^{\circ}$) and $92.4\pm0.5\,\%$ at 164\,nm ($45^{\circ}$).
Furthermore, the maximum reflectance at $65^{\circ}$ incidence
is still as high as $74\,\%$ at around 153\,nm.
Thus, this VUV-OC technique is useful in such a broadband VUV range,
and can be applicable to outcoupling 
e.g. the seventh harmonic at around 150\,nm 
generated from Yb-fiber combs at 1040--1060\,nm.
 
\subsection{Experimental results of intracavity HHG}
\label{intracavity_HHG}

\begin{figure*}[htb]
\resizebox{0.99\textwidth}{!}{%
  \includegraphics{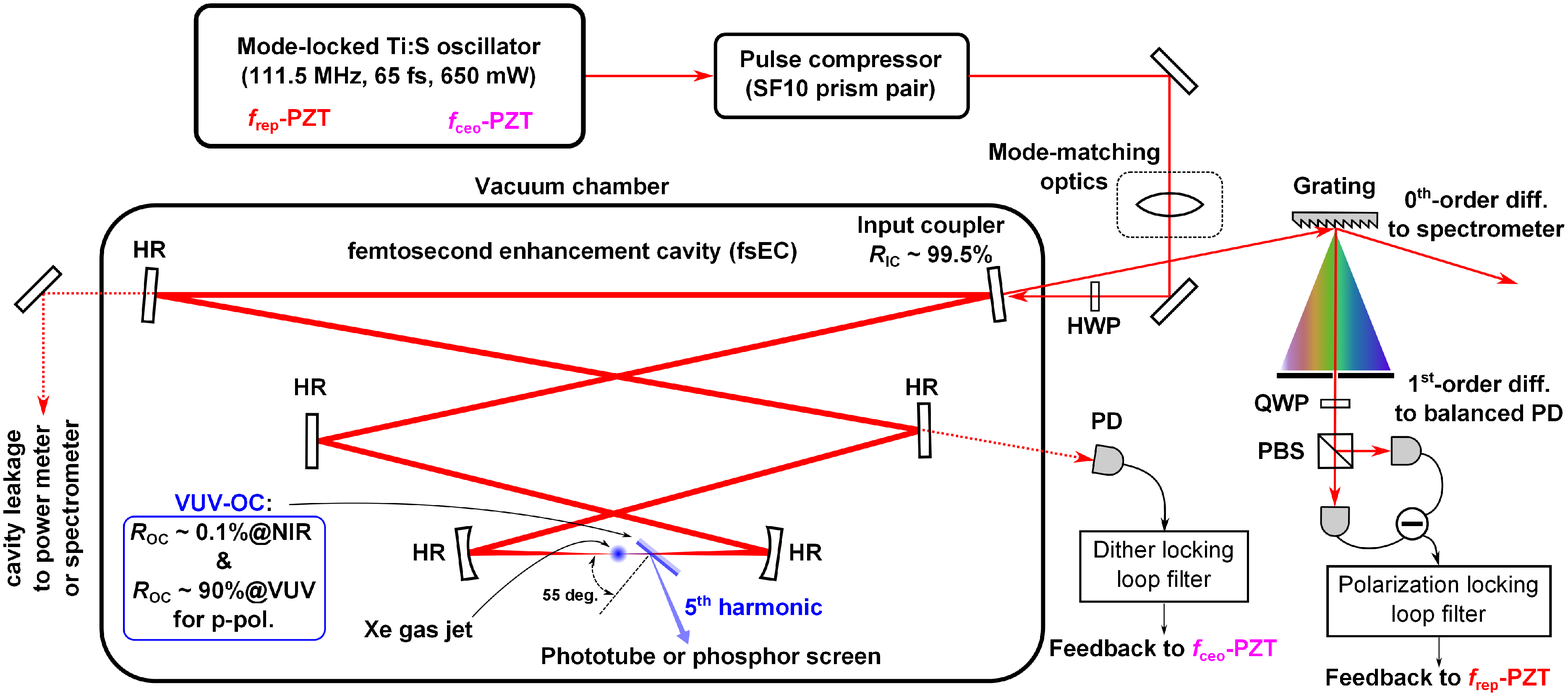}
}
\caption{Experimental setup for intracavity HHG using VUV-OC.
HWP: half-wave plate. QWP: quarter-wave plate.
PBS: polarization beam-splitter. PD: photo detector. 
HR: broadband high reflector with low group-delay dispersion. 
$f_{\mathrm{rep}}$-PZT and $f_{\mathrm{ceo}}$-PZT 
correspond to fast piezoelectric transducers 
on which cavity mirrors for the Ti:S oscillator are mounted and feedback-controlled.}
\label{exp_setup}       
\end{figure*}

We deployed our VUV-OC in our experimental intracavity HHG setup shown in Fig. \ref{exp_setup}.
A home-made Ti:S oscillator consists of a linear cavity including a prism pair. 
The center wavelength of fundamental fs combs is tuned to be 795\,nm. 
The repetition rate ($f_\mathrm{rep}$), spectral bandwidth, pulse width, and average power are 
111.5\,MHz, 20\,nm, 65\,fs and 650\,mW, respectively. 
After dispersion compensation by a pair of SF10 prisms, 
fs combs are guided into the fsEC placed inside a vacuum chamber. 
The fsEC consists of six broadband, low-GDD mirrors: 
an input coupler with a reflectance of $R_{\mathrm{IC}} \approx 99.5\,\%$, 
three planar high reflectors (HRs), 
and two concave HRs with radius of curvatures of 100\,mm. 
Although horizontal and vertical beam waist sizes between the two concave mirrors are slightly different  
because of aberration caused by the finite folding angle in bow-tie configuration, 
both of them are evaluated to be about $5\,\mu$m by 
beam propagation parameters measured outside the fsEC.
The reflection from the input coupler is guided to a diffraction grating. 
The zeroth-order diffraction from the grating is fiber-coupled and introduced into a spectrometer 
(USB4000, Ocean Optics) 
to evaluate reflection spectra from the fsEC.
Then the first-order diffraction is used for H\"{a}nsch-Couillaud locking technique \cite{HClocking80} 
(referred to as ``Polarization locking" in Fig. \ref{exp_setup}),
which adjusts $f_\mathrm{rep}$ of the Ti:S oscillator to that of the fsEC. 
The carrier-envelope-offset frequency $f_\mathrm{ceo}$ of the fs comb is controlled so that power inside the fsEC is maximum. 
Cavity leakage allows us to estimate the power as well as the spectrum of cavity field. 
The average circulating power inside the fsEC was estimated to be 120\,W (maximum),
which corresponds to the peak intensity of $2\times10^{13}$\,W/cm${}^2$.
Xe gas was used as nonlinear media and introduced to the beam waist
through a gas-jet nozzle (100\,$\mu$m diameter).
We observed visible plasma via ionization of the Xe gas.
The VUV-OC is placed at the Brewster angle of incidence right after the major focus of the fsEC. 
Generated VUV power was measured by a solar-blind phototube (R1187 \cite{R1187}, Hamamatsu). 
The phototube was calibrated at Hamamatsu, and the sensitivity to light at 160.8\,nm was measured to be 9\,mA/W. 

\begin{figure}
\resizebox{0.5\textwidth}{!}{%
  \includegraphics{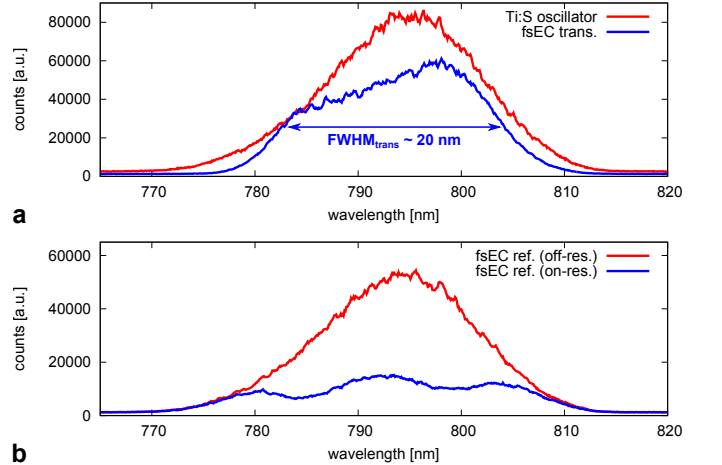}
}
\caption{
\textbf{(a)} Spectra of (red) Ti:S oscillator and (blue) fsEC transmission. 
\textbf{(b)} Reflection spectra when the fsEC is (red) off-resonant and (blue) on-resonant. 
}
\label{spectra}       
\end{figure}

Typical intensity spectra for the fundamental frequency combs are plotted in Fig. \ref{spectra}.
The red and blue solid curves in Fig. \ref{spectra}a, 
are spectra of the Ti:S oscillator and the fsEC transmission (leakage), respectively.
The bandwidth of transmitted spectrum is about 20\,nm, 
which we used to evaluate $\eta_{\rm{S}}$ as described in section \ref{feasibility_study}.
The blue (red) solid curve in Fig. \ref{spectra}b is a reflection spectrum from the fsEC
which is on-resonant (off-resonant). 
We roughly evaluate the frequency-averaged enhancement factor for the fsEC \cite{Ozawa08}:
$
\displaystyle
\mathcal{E} 
=
\frac{ 1- R_{\mathrm{IC}} }
{ 1 + \mathcal{R} -2\sqrt{ \mathcal{R} } },
$
obtained with neglecting a frequency-dependent round-trip phase term.
Here, 
$
\mathcal{R} = R_{\rm{IC}} \, (1-R_{\rm{OC}}^{\rm{NIR}}) \, R_{\rm{r}}, 
$
where $R_{\rm{r}}$ denotes residual reflectance 
in which we sum up reflectance for all the other mirrors except for the VUV-OC and the fsEC input coupler. 
We assume that VUV-OC and input coupler are lossless.
From the experimentally observed enhancement factor of 350,
we estimated $ R_{\rm{r}} $ to be $\approx 99.97\,\%$ for all the other five HR mirrors for fsEC.
Note, we neglect the GDD effect which might be caused by the VUV-OC.
Further precise measurement about frequency-resolved intracavity GDD and/or losses 
is possible by using several methods \cite{Thorpe05,Schliesser06,Hammond09}.

\begin{figure}
\resizebox{0.5\textwidth}{!}{%
  \includegraphics{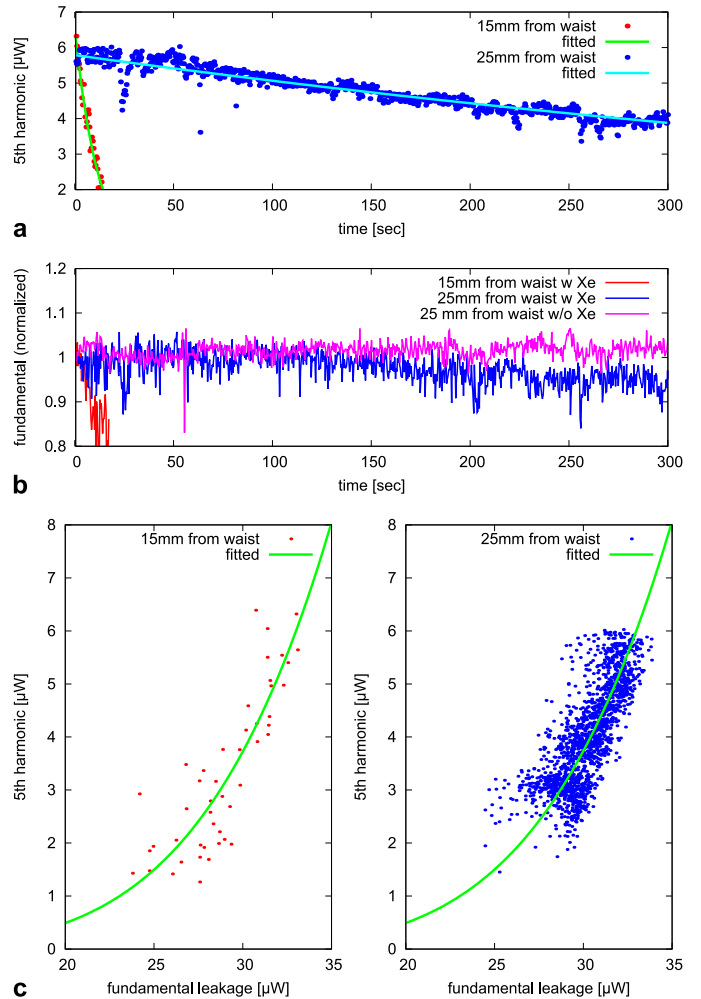}
}
\caption{
Time variation of \textbf{(a)} VUV power and \textbf{(b)} normalized fundamental power.
VUV-OC was placed at 15\,mm or 25\,mm away from the fsEC focus.
The fundamental and VUV power at the same VUV-OC position are synchronously sampled every $\sim0.3$\,s.
Time variation of a fundamental power in the absence of high harmonics 
is also shown as a magenta curve in \textbf{(b)}. 
\textbf{(c)} Yield of the fifth harmonic  
as a function of fundamental power in the two VUV-OC positions.
Green curves are obtained by linear least squares fitting using $ y = a x^5 $,
resulting in (left) $ a = (1.533 \pm 0.047) \times 10^{-7} $,
and (right) $ a = (1.538 \pm 0.006) \times 10^{-7} $.
Deviation of the harmonic yield at the same fundamental power 
is caused by slower response of the phototube compared to that of the power meter.
}
\label{VUV_decay_5thdep}       
\end{figure}

We then measured the time variation of average power in the VUV comb 
as shown in Fig. \ref{VUV_decay_5thdep}a.
We tested the two positions of the VUV-OC: (red) 15\,mm and (blue) 25\,mm away from the fsEC focus. 
As seen in the figure, the closer VUV-OC position resulted in the faster decay.
The decay constants are fitted by using $\displaystyle f(t) = A_0 \exp \left( -\frac{t-t_0}{\tau} \right)$,
and we obtained (green) $\tau_{\rm{15mm}} = 12.0\pm0.7$\,s, and (cyan) $\tau_{\rm{25mm}} = 750.0\pm9.3$\,s.
This decay is associated with the reduction of fundamental power inside the fsEC 
as can be seen from the Fig. \ref{VUV_decay_5thdep}b.
In contrast, we did not observe any power reduction (shown as the magenta curve)
without Xe gas, namely in the absence of high harmonics.
One of possible reasons for the degradation in the harmonic yield is 
VUV-induced IR absorption in our VUV-OC. 
VUV illumination may cause defects in the VUV-OC which can absorb intense IR beam. 
This may change GDD in the VUV-OC via accumulation of heat, which is not easily dissipated in vacuum. 
This degradation was not permanent 
and the harmonic yield was recovered after certain hours of break time. 
The similar reversible degradation at IR wavelength 
has been reported for oxide mirrors illuminated by UV radiation in free-electron-laser systems \cite{FELbook09}.

The yield of the fifth harmonic as a function of fundamental power 
is shown in Fig. \ref{VUV_decay_5thdep}c.
In both cases of (left) 15\,mm and (right) 25\,mm distance, 
data are fitted well to $y = ax^5$ as was predicted in section \ref{feasibility_study}. 
We measured the average power of fifth harmonic outcoupled from the fsEC to be $6.4\,\mu$W 
at maximum with a nozzle back pressure of about 1.4\,atm.
The HHG chamber was evacuated and the pressure inside the chamber was below 10\,mTorr when Xe gas was supplied. 
We previously reported 0.67\,$\mu$W \cite{Wakui11} using a diffraction-grating mirror as demonstrated in \cite{Yost08}.
Thus our VUV-OC outperforms our previous setup, and increased the final yield for nearly ten times.
Further optimization of HHG conditions \cite{Yost09,Carlson11,Allison11,Lee11,Hammond11} 
might yield power increase of the VUV combs.

\begin{figure}
\resizebox{0.5\textwidth}{!}{%
  \includegraphics{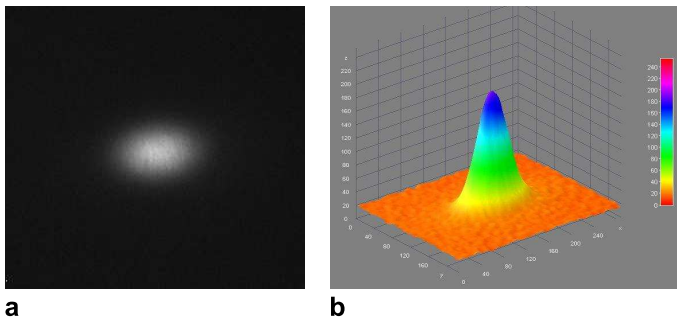}
}
\caption{\textbf{(a)} CCD image of fluorescence intensity 
excited by VUV radiation at a phosphor screen.
\textbf{(b)} 3D surface plot of the fluorescence intensity.}
\label{VUV_CCD_image}       
\end{figure}

\begin{figure}
\resizebox{0.5\textwidth}{!}{%
  \includegraphics{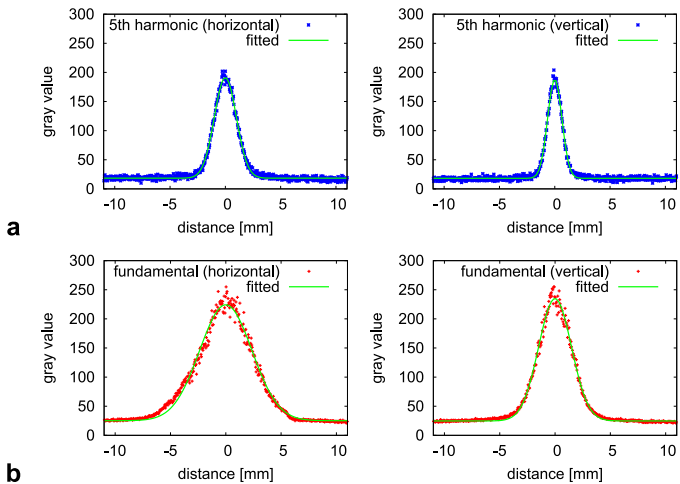}
}
\caption{
Experimental results of intensity distributions obtained from CCD images. 
\textbf{(a)} 
The left (right) figure is a horizontal (vertical) cross section of 
the fluorescence intensity distribution (shown in Fig. \ref{VUV_CCD_image}).
\textbf{(b)} 
The left (right) figure is a horizontal (vertical) cross section of
an intensity distribution for fundamental light scattered at the same phosphor screen.
In each case, the horizontal (vertical) axis is coincident with the major (minor) axis of an elliptic Gaussian.
}
\label{VUV_spatial_mode}       
\end{figure}

Finally, we measured a spatial-mode profile of VUV radiation outcoupled from the fsEC
via visible fluorescence emission from a phosphor screen covered with sodium salicylate 
(Scintillator window, McPherson). 
The fluorescence emission is peaked at 420\,nm \cite{Samson_book_67}. 
Because the VUV radiation is generated at the intracavity focus 
and reflected by the simple dichroic beamsplitter, 
the outcoupled VUV radiation should presumably be a clean Gaussian beam. 
This feature must be one of the indispensable prerequisites so as to realize tight focusing onto a trapped ion.
A snapshot of fluorescence intensity observed by a CCD camera (PL-B953U, PixeLINK) 
is shown in Fig. \ref{VUV_CCD_image}. 
The measured intensity distribution is a slightly asymmetric two-dimensional (2D) Gaussian.
%
%
We then fitted the fluorescence intensity and evaluated the spatial-mode profile of the VUV radiation, 
based on the fact that a quantum efficiency of sodium salycilate is high in the VUV region \cite{Allison64}
and is almost flat over that wavelength range \cite{Johnson51,Watanabe53}.
We also assume that the sodium salycilate layer was not saturated due to the faint ($\mu$W-level) average VUV power.
The fitting results by using a Gaussian function 
$ 
\displaystyle
f(x) = I \exp\left( -\frac{2(x-x_0)^2}{w^2} \right) + I_0,
$
are shown in Fig. \ref{VUV_spatial_mode}a.
We obtained 
(left) $w_{\mathrm{h}}^{\mathrm{5th}} = 2.00 \pm 0.01$\,mm 
and
(right) $w_{\mathrm{v}}^{\mathrm{5th}} = 1.29 \pm 0.01$\,mm 
as a beam radius at the phosphor screen,
which is placed at about 140\,mm away from the fsEC focus.
Thus a beam diameter of the fifth harmonic on the VUV-OC 
is estimated to be about 1\,mm in the case of 25\,mm distance.
We also evaluated a beam radius of fundamental light 
by the same CCD camera observing the fundamental light 
scattered at the same phosphor screen.
The fitting results are shown in Fig. \ref{VUV_spatial_mode}b, 
resulting in 
(left) $w_{\mathrm{h}}^{\mathrm{F}} = 4.72 \pm 0.02$\,mm 
and
(right) $w_{\mathrm{v}}^{\mathrm{F}} = 2.97 \pm 0.01$\,mm.
In such a way 
we obtained a beam divergence ratio of the fifth harmonic compared to the fundamental as 
$w_{\mathrm{h}}^{\mathrm{5th}} 
/ w_{\mathrm{h}}^{\mathrm{F}}
\approx 0.42$
and
$w_{\mathrm{v}}^{\mathrm{5th}} 
/ w_{\mathrm{v}}^{\mathrm{F}}
\approx 0.43$.
According to the theoretical prediction \cite{Pupeza13},
the ratio is expected to be $1/\sqrt{5} \approx 0.45$,
which is in good agreement with the above experimental observation.
Thus we confirmed that only the fifth harmonic was successfully outcoupled from the fsEC
as the Gaussian beam using our VUV-OC.

\section{Conclusion}
\label{conclusion}

In conclusion, we have reported a feasibility study on detection of a single trapped In${}^+$ at 159\,nm VUV wavelength. 
We estimated a fluorescence photon counting rate per the average VUV power to be $n_0/P_0 \approx 87$\,cps/$\mu$W 
in a realistic experimental condition, in which the ion is irradiated by a quasi-CW VUV beam generated via intracavity HHG. 
This suggests that an average VUV power of microwatt level suffices for detecting an In${}^+$ ion. 
Toward this goal we have built an intracavity HHG setup for generating VUV radiation using a novel efficient output coupler (VUV-OC). 
Reflectance of the VUV-OC at an incident angle of $55^{\circ}$ was measured to be $86.9\pm0.4\%$ at 159\,nm, and was achieved $>80\,\%$ from 155 to 161.5\,nm.
Our HHG setup with the VUV-OC has provided an output of $6.4\,\mu$W VUV combs around 159\,nm with an unexpected exponential decay in power. 
This power level will yield more than 500\,cps detection of fluorescent photons, 
indicating that quantum state discrimination of the ion is feasible with our present setup, 
provided that the power decay problem is properly addressed.

With minor modifications and further improvements, 
our HHG setup can be deployed to a wider range of applications to optical clocks based on ions with two outer elections. 
A Ti:S oscillator centered at 835\,nm can be used for generating VUV radiation at 167\,nm as the fifth harmonic in a setup similar to ours, 
and our VUV-OC approach provides efficient output coupling also at this wavelength as shown in Fig. \ref{reflectance}. 
The 167\,nm radiation generated in this way can be used for probing the ${}^1\rm{S}_0$--${}^1 \rm{P}_1$ transition of Al${}^+$. 
Boosting of the fundamental power up to 1\,W is possible with commercially available fs combs, 
and it gives a peak intensity of $3 \times 10^{13}$\,W/cm${}^2$ in our setup. 
In this perturbative regime \cite{Gohle05} the HHG power scaling can be roughly estimated based on our experimental data shown in Fig. \ref{VUV_decay_5thdep}c. 
An average VUV comb power of 50\,$\mu$W is estimated for the 1\,W input, 
corresponding to a detection rate of more than 4000 cps from an In${}^+$. 
This counting rate could dramatically reduce the detection time. 
Further increase of the fundamental fs comb power can be achieved using an injection locking amplifier \cite{Paul08}, 
but the scaling of the VUV power does not follow the perturbative approximation any more. 
Nevertheless increase in the VUV power is expected, as has been reported in ref \cite{Lee11}, 
in which XUV radiation of 77\,$\mu$W at 72\,nm has been generated as the 11th harmonic of the amplified Ti:S oscillator. 
In such power level of tens of microwatt, instead of focusing the quasi-CW beam tightly to a single ion, 
the beam could be loosely focused to illuminate a chain of multiple ions simultaneously along the trapping axis. 
This might provide a novel detection method for multi-ion optical clocks.

\section*{Acknowledgments}
The authors thank 
H. Hachisu, A. Yamaguchi, Y. Li, S. Nagano, H. Ishijima, M. Kumagai, M. Kajita, Y. Hanado and M. Sasaki 
for discussions and support,
and E. Sasaki for technical support.
Part of this work was supported by the Use-of-UVSOR Facility Program (BL5B, No. 25-568) 
of the Institute for Molecular Science.

%
%

\end{document}